\documentclass[10pt,twocolumn]{IEEEtran}
\usepackage{amsbsy}
\usepackage{amsmath}
\usepackage{times}
\input epsf
\title{Minimum Mean-Squared Error Iterative Successive
Parallel Arbitrated Decision Feedback Detectors for DS-CDMA
Systems}

\author{Rodrigo C. de Lamare and Raimundo Sampaio-Neto  \\
\thanks{R. C. de Lamare is with the Communications Research Group, Department of Electronics, University of York,
York Y010 5DD, United Kingdom and R. Sampaio-Neto is with
CETUC/PUC-RIO, 22453-900, Rio de Janeiro - Brazil Phone:
+55-21-31141701. E-mails: rcdl500@ohm.york.ac.uk,
raimundo@cetuc.puc-rio.br} }

\begin{document}
\maketitle

\begin{abstract}
{In this paper we propose minimum mean squared error (MMSE)
iterative successive parallel arbitrated decision feedback (DF)
receivers for direct sequence code division multiple access
(DS-CDMA) systems. We describe the MMSE design criterion for DF
multiuser detectors along with successive, parallel and iterative
interference cancellation structures. A novel efficient DF
structure that employs successive cancellation with parallel
arbitrated branches and a near-optimal low complexity user
ordering algorithm are presented. The proposed DF receiver
structure and the ordering algorithm are then combined with
iterative cascaded DF stages for mitigating the deleterious
effects of error propagation for convolutionally encoded systems
with both Viterbi and turbo decoding as well as for uncoded
schemes. We mathematically study the relations between the MMSE
achieved by the analyzed DF structures, including the novel
scheme, with imperfect and perfect feedback. Simulation results
for an uplink scenario assess the new iterative DF detectors
against linear receivers and evaluate the effects of error
propagation of the new cancellation methods against existing
ones.}
\end{abstract}

\section{Introduction}

Multiuser detection has been proposed as a means to suppress
multi-access interference (MAI), increasing the capacity and the
performance of CDMA systems \cite{verdu}. The optimal multiuser
detector of Verdu \cite{verdu86} suffers from exponential
complexity and requires the knowledge of timing, amplitude and
signature sequences. This fact has motivated the development of
various sub-optimal strategies: the linear \cite{lupas} and
decision feedback (DF) \cite{falconer} receivers, the successive
interference canceller \cite{patel} and the multistage detector
\cite{varanasi}. Recently, Verdu and Shamai \cite{shamai} and
Rapajic \cite{rapajic} {\it et al.} have investigated the
information theoretic trade-off between the spectral and power
efficiency of linear and non-linear multiuser detectors in
synchronous AWGN channels. These works have shown that given a
sufficient signal to noise ratio and for high loads (the ratio of
users to processing gain close to one), DF detection has a
substantially higher spectral efficiency than linear detection.
For uplink scenarios, DF structures, which are relatively simple
and perform linear interference suppression followed by
interference cancellation, provide substantial gains over linear
detection.

Minimum mean squared error (MMSE) multiuser detectors usually show
good performance and have simple adaptive implementation. In
particular, when used with short or repeated spreading sequences
the MMSE design criterion leads to adaptive versions which only
require a training sequence for estimating the receiver
parameters. Previous work on DF detectors examined successive
interference cancellation \cite{duel,varanasi2,varanasi3},
parallel interference cancellation
\cite{woodward1,woodward2,woodward3} and multistage or iterative
DF detectors \cite{woodward2,woodward3}. The DF detector with
successive interference cancellation (S-DF) is optimal, in the
sense that it achieves the sum capacity of the the synchronous
AWGN channel \cite{varanasi2}. The S-DF scheme is capable of
alleviating the effects of error propagation despite it generally
leads to non uniform performance over the users. In particular,
the user ordering plays an important role in the performance of
S-DF detectors. Studies on decorrelator DF detectors with optimal
user ordering have been reported in \cite{varanasi3} for imperfect
feedback and in \cite{luo} for perfect feedback. The problem with
the optimal ordering algorithms in \cite{varanasi3,luo} is that
they represent a very high computational burden for practical
receiver design. Conversely, the DF receiver with parallel
interference cancellation (P-DF)
\cite{woodward1,woodward2,woodward3} satisfies the uplink
requirements, namely, cancellation of intracell interference and
suppression of the remaining other-cell interference, and
provides, in general, uniform performance over the user population
even though it is more sensitive to error propagation. The
multistage or iterative DF schemes presented in
\cite{woodward2,woodward3} are based on the combination of S-DF
and P-DF schemes in multiple stages in order to refine the symbol
estimates, resulting in improved performance over conventional
S-DF, P-DF and mitigation of error propagation.

In this work, we propose the design of MMSE DF detectors that
employ a novel successive parallel arbitrated DF (SPA-DF)
structure based on the generation of parallel arbitrated branches.
The motivation for the novel DF structures is to mitigate the
effects of error propagation often found in P-DF structures
\cite{woodward1,woodward2,woodward3}. The basic idea is to improve
the S-DF structure using different orders of cancellation and then
select the most likely estimate. A near-optimal user ordering
algorithm is described for the new SPA-DF detector structure and
is compared to the optimal user ordering algorithm, which requires
the evaluation of $K!$ different cancellation orders. The results
in terms of performance show that the SPA-DF structure with the
suboptimal ordering algorithm can achieve a performance very close
to that of the S-DF with optimal ordering. Furthermore, the new
SPA-DF scheme is combined with iterative cascaded DF stages, where
the subsequent stage uses S-DF, P-DF or the new SPA-DF system to
refine the symbol estimates of the users and combat the effects of
error propagation. The performance of the proposed SPA-DF scheme
and the sub-optimal ordering algorithm and their combinations with
other schemes in a multistage detection structure is investigated
for both uncoded and convolutionally encoded systems with Viterbi
and turbo decoding.

This paper is structured as follows. Section II briefly describes
the DS-CDMA system model. The MMSE decision feedback receiver
filters are described in Section III. Sections IV is devoted to
the novel SPA-DF scheme, the near-optimal user ordering algorithm
and the combination of the SPA-DF detector with iterative cascaded
DF stages and Section V details the proposed SPA-DF receiver for
convolutionally coded systems with Viterbi and turbo decoding.
Section VI presents and discusses the simulation results and
Section VII draws the concluding remarks of this paper.

\section{DS-CDMA system model}

Let us consider the uplink of a symbol synchronous binary
phase-shift keying (BPSK) DS-CDMA system with $K$ users, $N$ chips
per symbol and $L_{p}$ propagation paths. It should be remarked
that a synchronous model is assumed for simplicity, although it
captures most of the features of more realistic asynchronous
models with small to moderate delay spreads. The baseband signal
transmitted by the $k$-th active user to the base station is given
by {
\begin{equation}
x_{k}(t)=A_{k}\sum_{i=-\infty}^{\infty}b_{k}(i)s_{k}(t-iT)\end{equation}
} where $b_{k}(i) \in \{\pm1\}$ denotes the $i$-th symbol for user
$k$, the real valued spreading waveform and the amplitude
associated with user $k$ are $s_{k}(t)$ and $A_{k}$, respectively.
The spreading waveforms are expressed by { $s_{k}(t) =
\sum_{i=1}^{N}a_{k}(i)\phi(t-iT_{c})$}, where { $a_{k}(i)\in
\{\pm1/\sqrt{N} \}$}, { $\phi(t)$} is the chip waveform, $T_{c}$
is the chip duration and $N=T/T_{c}$ is the processing gain.
Assuming that the receiver is synchronised with the main path, the
coherently demodulated composite received signal is
\begin{equation}
r(t)= \sum_{k=1}^{K}\sum_{l=0}^{L_{p}-1}
h_{k,l}(t)x_{k}(t-\tau_{k,l})+n(t)
\end{equation}
where $h_{k,l}(t)$ and $\tau_{k,l}$ are, respectively, the channel
coefficient and the delay associated with the $l$-th path and the
$k$-th user. Assuming that $\tau_{k,l} = lT_{c}$, the channel is
constant during each symbol interval, the spreading codes are
repeated from symbol to symbol and the receiver is synchronized
with the main path, the received signal $ r(t)$ after filtering by
a chip-pulse matched filter and sampled at chip rate yields the
$M$-dimensional received vector
\begin{equation}
\begin{split}
{\bf r}(i) = & \sum_{k=1}^{K}A_{k}b_{k}(i){\bf C}_{k}{\bf
h}_{k}(i)  + A_{k}b_{k}(i-1)\bar{\bf C}_{k}{\bf h}_{k}(i-1) \\ & +
A_{k}b_{k}(i+1)\breve{\bf C}_{k}{\bf
h}_{k}(i+1) + {\bf n}(i) \\
 = & \sum_{k=1}^{K}\Big(A_{k}b_{k}(i){\bf p}_{k}(i) + \boldsymbol{\eta}_{k}(i)\Big) + {\bf n}(i)
\end{split}
\end{equation}
where $M=N+L_{p}-1$, ${\bf n}(i) = [n_{1}(i)
~\ldots~n_{M}(i)]^{T}$ is the complex gaussian noise vector with
$E[{\bf n}(i){\bf n}^{H}(i)] = \sigma^{2}{\bf I}$, $(.)^{T}$ and
$(.)^{H}$ denote transpose and Hermitian transpose, respectively,
$E[.]$ stands for ensemble average, $b_{k}(i) \in \{\pm1+j0\}$ is
the symbol for user $k$,  the amplitude of user $k$ is $A_{k}$,
the user $k$ channel vector is ${\bf h}_{k}(i) = [h_{k,0}(i)
\ldots h_{k,L_{p}-1}(i)]^{T}$ with $h_{k,l}(i) = h_{k,l}(iT_{c})$
for $l=0,\ldots,L_{p}-1$, the ISI is given by
$\boldsymbol{\eta}_{k}(i) = A_{k}b_{k}(i-1)\bar{\bf C}_{k}{\bf
h}_{k}(i-1) + A_{k}b_{k}(i+1)\breve{\bf C}_{k}{\bf h}_{k}(i+1)$
and assumes that the channel order is not greater than $N$, i.e.
$L_p-1 \leq N$, ${\bf s}_{k} = [a_{k}(1) \ldots a_{k}(N)]^{T}$ is
the signature sequence for user $k$ and ${\bf p}_{k}(i) = {\bf
C}_{k}{\bf h}_{k}(i)$ is the effective signature sequence for user
$k$, the $M \times L_{p}$ convolution matrix ${\bf C}_{k}$
contains one-chip shifted versions of ${\bf s}_{k}$ and the $M
\times L_{p}$ matrices $\bar{\bf C}_{k}$ and $\breve{\bf C}_{k}$
with segments of ${\bf s}_{k}$ have the following structure {
$${\bf C}_{k} = \left[\hspace*{-0.5em}\begin{array}{c c c c }
a_{k}(1) & 0 & \ldots & {0} \\
\vdots & a_{k}(1) & \ddots & \vdots  \\
a_{k}(N) & \vdots & \ddots & 0 \\
0 & a_{k}(N) & \ddots & a_{k}(1)  \\
\vdots & \vdots &  \ddots & \vdots  \\
0 & 0& \ddots & a_{k}(N)  \\
 \end{array}\hspace*{-0.5em}\right], $$ $$ \bar{\bf C}_{k} =
\left[\hspace*{-0.5em}\begin{array}{c c c c}
0 & a_k(N) & \ldots & a_k(N-L_p+1) \\
\vdots &  0 &  \ddots & \vdots \\
0 &  \vdots & \ddots &  a_k(N) \\
\vdots & 0 & \ddots & 0  \\
0 & \vdots & \ddots & 0  \\
0 & 0 & \ldots & 0  \\
 \end{array}\hspace*{-0.5em}\right], $$ $$
\breve{\bf C}_{k} = \left[\hspace*{-0.5em}\begin{array}{c c c c}
0 & \ldots & 0 &  {0} \\
\vdots &  \ldots & \vdots &  \vdots  \\
0 & \ldots & 0 &  0 \\
a_{k}(1) & \ddots & 0  & 0  \\
\vdots & \ddots & \vdots &   \vdots  \\
a_{k}(L_p-1) & \ldots & a_{k}(1)  & 0  \\
 \end{array}\hspace*{-0.5em}\right].$$}

The MAI comes from the non-orthogonality between the received
signature sequences, whereas the ISI span $L_{s}$ depends on the
length of the channel response, which is related to the length of
the chip sequence. For { $L_{p}=1,~ L_{s}=1$} (no ISI), for {
$1<L_{p}\leq N, L_{s}=2$}, for { $N <L_{p}\leq 2N, L_{s}=3$}.

\section{MMSE Decision Feedback Receivers}

Let us describe in this section the design of synchronous MMSE
decision feedback detectors. The input to the hard decision device
corresponding to the $i$th symbol is ${\bf z}(i) = {\bf
W}^{H}(i){\bf r}(i) - {\bf F}^{H}(i)\hat{\bf b}(i)$, where the
input ${\bf z}(i) = [z_{1}(i) ~\ldots ~z_{K}(i)]^{T}$, ${\bf W}(i)
= [{\bf w}_{1}~ \ldots~{\bf w}_{K}]$ is $M \times K$ the
feedforward matrix, $\hat{\bf b}(i)=[b_{1}(i)
~\ldots~b_{K}(i)]^{T}$ is the $K \times 1$ vector of estimated
symbols, which are fed back through the $K \times K$ feedback
matrix ${\bf F}(i)=[ {\bf f}_{1}(i)~\ldots~{\bf f}_{K}(i)]$.
Generally, the DF receiver design is equivalent to determining for
user $k$ a feedforward filter { ${\bf w}_{k}(i)$} with { $M$}
elements and a feedback one { ${\bf f}_{k}(i)$} with $K$ elements
that provide an estimate of the desired symbol:
\begin{equation}
z_{k}(i) = {\bf w}_{k}^{H}(i){\bf r}(i)-{\bf f}_{k}^{H}(i)\hat{\bf
b}(i)~, ~~~~~~k=1,2,\ldots,K
\end{equation}
where $\hat{\bf b}(i)={\rm sgn}[\Re({\bf W}^{H}{\bf r}(i))]$ is
the vector with initial decisions provided by the linear section,
${\bf w}_{k}$ and ${\bf f}_{k}$ are optimized by the MMSE
criterion. In particular, the feedback filter ${\bf f}_{k}(i)$ of
user $k$ has a number of non-zero coefficients corresponding to
the available number of feedback connections for each type of
cancellation structure. The final detected symbol is:
\begin{equation}
 \hat{b}_{k}^{f}(i) = {\rm sgn}\Big(\Re\Big[z_{k}(i)\Big]\Big) = {\rm sgn}\Big(\Re\Big[{\bf w}_{k}^{H}(i){\bf
r}(i)-{\bf f}_{k}^{H}(i)\hat{\bf b}(i)\Big]\Big)
\end{equation}
where the operator $(.)^{H}$ denotes Hermitian transpose, $\Re(.)$
selects the real part and ${\rm sgn}(.)$ is the signum function.

To describe the optimal MMSE filters we will initially assume
perfect feedback, that is $\hat{\bf b}= {\bf b}$, and then will
consider a more general framework. Consider the following cost
function:
\begin{equation}
J_{MSE} = E\Big[ |b_{k}(i) - {\bf w}_{k}^{H}{\bf r}(i) + {\bf
f}_{k}^{H}{\bf b}(i)|^{2} \Big]
\end{equation}

Let us divide the users into two sets, similarly to
\cite{woodward2}
\begin{equation}
D = \{j:{\hat b}_{j}  ~ is ~fed ~back ~\}
\end{equation}
\begin{equation}
U = \{j: j \notin D \}
\end{equation}
where the two sets $D$ and $U$ correspond to detected and
undetected users, respectively. Let us also define the matrices of
effective spreading sequences ${\bf P} = [{\bf p}_{1} ~\ldots ~
{\bf p}_{K}]$, ${\bf P}_{D} = [{\bf p}_{1} ~\ldots ~ {\bf p}_{D}]$
and ${\bf P}_{U} = [{\bf p}_{1} ~\ldots ~ {\bf p}_{U}]$. The
minimization of the cost function in (6) with respect to the
filters ${\bf w}_{k}$ and ${\bf f}_{k}$ yields:
\begin{equation}
{\bf w}_{k} = {\bf R}_{U}^{-1}{\bf p}_{k}
\end{equation}
\begin{equation}
{\bf f}_{k} = {\bf P}_{D}^{H}{\bf w}_{k}
\end{equation}
where the associated covariance matrices are ${\bf R} = E[{\bf
r}(i){\bf r}^{H}(i)] = {\bf P}{\bf P}^{H} + \sigma^{2}{\bf I}$,
${\bf R}_{U} = {\bf P}_{U}{\bf P}^{H}_{U} + \sigma^{2}{\bf I} =
{\bf R} - {\bf P}_{D}{\bf P}_{D}^{H}$. Thus, assuming perfect
feedback and that user $k$ is the desired one, the associated MMSE
for the DF receiver is given by:
\begin{equation}
J_{MMSE} = \sigma^{2}_{b} - {\bf p}_{k}^{H}{\bf R}_{U}^{-1}{\bf
p}_{k}
\end{equation}
where $\sigma^{2}_{b}=E[|b_{k}^{2}(i)|]$. The result in (11) means
that in the absence of error propagation, the MAI in set $D$ is
eliminated and user $k$ is only affected by interferers in set
$U$.

For the successive interference cancellation DF (S-DF) detector ,
we have for user $k$
\begin{equation}
D= \{1, ~\ldots~, k-1\}, ~~~~~ U= \{k, ~\ldots~ , K\}
\end{equation}
where the filter matrix ${\bf F}(i)$ is strictly upper triangular.
The S-DF structure is optimal in the sense of that it achieves the
sum capacity of the synchronous CDMA channel with AWGN
\cite{varanasi2}. In addition, the S-DF scheme is less affected by
error propagation although it generally does not provide uniform
performance over the user population. In order to design the S-DF
receivers and satisfy the constraints of the S-DF structure, the
designer must obtain the vector with initial decisions $\hat{\bf
b}(i)={\rm sgn}[\Re({\bf W}^{H}(i){\bf r}(i))]$ and then resort to
the following cancellation approach. The non-zero part of the
filter ${\bf f}_{k}$ corresponds to the number of used feedback
connections and to the users to be cancelled. For the S-DF, the
number of feedback elements and their associated number of
non-zero filter coefficients in ${\bf f}_{k}$ (where $k$ goes from
the second detected user to the last one) range from $1$ to $K-1$.

The parallel interference cancellation DF (P-DF) \cite{woodward2}
receiver can offer uniform performance over the users but it
suffers from error propagation. For the P-DF in a single cell, we
have \cite{woodward2}
\begin{equation}
D= \{1, ~\ldots~, k-1\, k+1, ~\ldots, K \}, ~~~~~ U= \{k \}
\end{equation}
\begin{equation}
{\bf w}_{k} = {\bf R}_{U}^{-1}{\bf p}_{k} = \frac{{\bf
p}_{k}}{A_{k}^{2} + \sigma^{2}}
\end{equation}
The MMSE associated with the P-DF system is obtained by
substituting ${\bf R}_{U} = {\bf R} - {\bf P}_{D}{\bf P}_{D}^{H}$
into (9), which yields:
\begin{equation}
J_{MMSE} = \sigma^{2}_{b} - {\bf p}_{k}^{H}({\bf p}_{k}{\bf
p}_{k}^{H} + \sigma^{2}{\bf I})^{-1}{\bf p}_{k} =
\frac{\sigma^{2}}{A_{k}^{2} + \sigma^{2} }
\end{equation}
where for P-DF  ${\bf F}(i)$ is full and constrained to have zeros
along the diagonal to avoid cancelling the desired symbols. In
order to design P-DF receivers and satisfy their constraints, the
designer must obtain the vector with initial decisions $\hat{\bf
b}(i)={\rm sgn}[\Re({\bf W}^{H}(i){\bf r}(i))]$ and then resort to
the following cancellation approach. The non-zero part of the
filter ${\bf f}_{k}$ corresponds to the number of used feedback
connections and to the users to be cancelled. For the P-DF, the
feedback connections used and their associated number of non-zero
filter coefficients in ${\bf f}_{k}$ are equal to $K-1$ for all
users and the matrix ${\bf F}(i)$ has zeros on the main diagonal
to avoid cancelling the desired symbols.

Now let us consider a more general framework, where the feedback
is not perfect. The minimization of the cost function in (4) with
respect to ${\bf w}_{k}$ and ${\bf f}_{k}$ leads to the following
filter expressions:
\begin{equation}
{\bf w}_{k} = {\bf R}^{-1}({\bf p}_{k} + {\bf B}{\bf f}_{k})
\end{equation}
\begin{equation}
{\bf f}_{k} = (E[\hat{\bf b}\hat{\bf b}^{H}])^{-1} {\bf B}^{H}{\bf
w}_{k} \approx {\bf B}^{H}{\bf w}_{k}
\end{equation}
where $E[\hat{\bf b}\hat{\bf b}^{H}] \approx {\bf I}$ for small
error rates and ${\bf B}=E[{\bf r}(i)\hat{\bf b}^{H}(i)]$. The
associated MMSE for DF receivers subject to $E[\hat{\bf b}\hat{\bf
b}^{H}] \approx {\bf I}$ and imperfect feedback is approximately
given by
\begin{equation}
J_{MMSE} \approx \sigma^{2}_{b} - {\bf p}_{k}^{H}{\bf R}^{-1}{\bf
p}_{k} - {\bf p}_{k}^{H}{\bf R}^{-1}{\bf B}{\bf f}_{k}
\end{equation}
In Appendix I we show that the expression in (18) equals (11)
under perfect feedback, and provide several other relationships
between DF structure with and without perfect feedback. Note that
the MMSE associated with DF receivers that are subject to
imperfect feedback depends on the matrix ${\bf B}=E[{\bf
r}\hat{\bf b}^{H}]$, that under perfect feedback equals ${\bf
P}_{D}$, and the feedback filter ${\bf f}_{k}$ or set of filters
${\bf F}$. Specifically, if we choose a given structure for ${\bf
F}$ this approach will lead to different methods of interference
cancellation and performance improvements for the DF detector as
compared to linear detection. The motivation for our work is to
investigate alternative methods of finding structures for ${\bf
F}$ that provide enhanced performance.

\section{Successive Parallel Arbitrated DF and Iterative Detection}

In this section, we present a novel interference cancellation
structure and describe a low complexity near-optimal ordering
algorithm that employs different orders of cancellation and then
selects the most likely symbol estimate. The proposed ordering
algorithm is compared with the optimal user ordering algorithm,
which requires the evaluation of $K!$ different cancellation
orders and turns out to be too complex for practical use. The new
receiver structure, denoted successive parallel arbitrated DF
(SPA-DF) detection, is then combined with iterative cascaded DF
stages \cite{woodward2,woodward3} to further refine the symbol
estimates. The motivation for the novel DF structures is to
mitigate the effects of error propagation often found in P-DF
structures \cite{woodward2,woodward3}, that are of great interest
for uplink scenarios due to its capability of providing uniform
performance over the users.

\subsection{Successive Parallel Arbitrated DF Detection}

The idea of parallel arbitration is to employ successive
interference cancellation (SIC) to rapidly converge to a local
maximum of the likelihood function and, by running parallel
branches of SIC with different orders of cancellation, one can
arrive at sufficiently different local maxima \cite{barriac}. The
goal of the new scheme, whose block diagram is shown in Fig. 1, is
to improve performance using parallel searches and to select the
most likely symbol estimate. The idea of the ordering algorithm is
to employ SIC for different branches based on the power of the
users to rapidly converge to a local maximum of the likelihood
function and, on the basis of the euclidean distance, our approach
selects the most likely estimate. In order to obtain the benefits
of parallel search, the candidates should be arbitrated, yielding
different estimates of a symbol. The estimate of a symbol that has
the highest likelihood is then selected at the output. Unlike the
work of Barriac and Madhow \cite{barriac} that employed matched
filters as the starting point, we adopt MMSE DF receivers as the
initial condition and the euclidean distance for selecting the
most likely symbol. The concept of parallel arbitration is thus
incorporated into a DF detector structure, that applies linear
interference suppression followed by SIC and yields improved
starting points as compared to matched filters. Note that our
approach does not require signal reconstruction as the PASIC in
\cite{barriac} because the MMSE filters automatically compute the
coefficients for interference cancellation.

\begin{figure}[!htb]
\begin{center}
\def\epsfsize#1#2{1\columnwidth}
\epsfbox{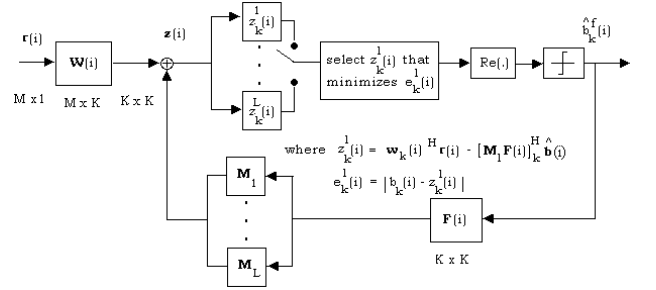} \caption{Block diagram of the proposed SPA-DF
receiver. }
\end{center}
\end{figure}

Following the schematic of Fig. 1, the user $k$ output of the
parallel branch $l$ ($l=1,~ \ldots,~ L$) for the SPA-DF receiver
structure is given by:
\begin{equation}
z_{k}^{l}(i) = {\bf w}^{H}_{k}(i){\bf r}(i) - [{\bf M}_{l}{\bf
F}]_{k}^{H}\hat{\bf b}(i)
\end{equation}
where $\hat{\bf b}(i)={\rm sgn}[\Re({\bf W}^{H}{\bf r}(i))]$ and
the matrices ${\bf M}_{l}$ are permutated square identity (${\bf
I}_{K}$) matrices with dimension $K$ whose structures for an
$L=4$-branch SPA-DF scheme are given by:
\begin{equation}
{\bf
M}_{1}= {\bf I}_{K},~ {\bf M}_{2}=\left[\begin{array}{cc}
{\bf 0}_{K/4,3K/4} & {\bf I}_{3K/4} \\
{\mathbf{I}}_{K/4} & {\bf 0}_{K/4,3K/4}
\end{array}\right], ~ $$ $$
{\bf M}_{3}=\left[\begin{array}{cc}
{\bf 0}_{K/2} & {\bf I}_{K/2} \\
{\bf I}_{K/2} & {\bf 0}_{K/2}
\end{array}\right],~
{\bf M}_{4}=\left[\begin{array}{ccc}
 0 & \ldots & 1 \\
\vdots & . \cdot {\Large }^{\Large .} &  \vdots \\
1 & \ldots  & 0
\end{array}\right]
\end{equation}
where ${\bf 0}_{m,n}$ denotes an $m \times n$-dimensional matrix
full of zeros and the structures of the matrices $M_{l}$
correspond to phase shifts regarding the cancellation order of the
users. The purpose of the matrices in (20) is to change the order
of cancellation. When ${\bf M}={\bf I}$ the order of cancellation
is a simple successive cancellation (S-DF) based upon the user
powers (the same as \cite{duel,varanasi2}). Specifically, the
above matrices perform the cancellation with the following order
with respect to user powers: $M_{1}$ with indices $1, \ldots, K$;
$M_{2}$ with indices $K/4,K/4+1, \ldots, K,1,\ldots,K/4-1$;$M_{3}$
with indices $K/2,K/2+1, \ldots, K,1,\ldots,K/2-1$; $M_{4}$ with
$K, \ldots, 1$ (reverse order). The proposed ordering algorithm
shifts the ordering of the users according to $K/B$, where $B$ is
the number of parallel branches. The rationale for this approach
is to shift the ordering and attempt to benefit a given user or
group of users for each decoding branch. Following this approach,
a user that for a given ordering appears to be in an unfavorable
position can benefit in other parallel branches by being detected
in a more favorable situation. For more branches, additional phase
shifts are applied with respect to user cancellation ordering.
Note that different update orders were tested although they did
not result in performance improvements.

The final output $\hat{b}_{k}^{f}(i)$ of the SPA-DF detector
chooses the best estimate of the $L$ candidates for each symbol
interval $i$ as described by:
\begin{equation}
\hat{b}_{k}^{(f)}(i) = {\rm sgn}\Big[ \Re \Big( \arg \min_{1 \leq
l \leq L} e_{k}^{l}(i) \Big) \Big]
\end{equation}
where the best estimate is the value $z_{k}^{l}(i)$ that minimizes
$e_{k}^{l}(i)=|b_{k}(i) - z_{k}^{l}(i)|$ and
$\hat{b}_{k}^{(f)}(i)$ forms the vector of final decisions
$\hat{\bf b}_{k}^{(f)}(i) = [\hat{b}_{1}^{(f)}(i)~\ldots
~\hat{b}_{K}^{(f)}(i)]^{T}$. The number of parallel branches $L$
that yield detection candidates is a parameter that must be chosen
by the designer. In this context, the optimal ordering algorithm
conducts an exhaustive search and is given by
\begin{equation}\hat{b}_{k}^{(f)}(i) = {\rm sgn}\Big[ \Re
\Big( \arg \min_{1 \leq l \leq K!} e_{k}^{l}(i) \Big)
\Big]
\end{equation}
where the number of candidates is $L=K!$ and is clearly very
complex for practical systems. Our studies indicate that $L=4$
achieves most of the gains of the new structure and offers a good
trade-off between performance and complexity. The SPA-DF system
employs the same filters, namely ${\bf W}$ and ${\bf F}$, of the
traditional S-DF structure and requires additional arithmetic
operations to compute the parallel arbitrated candidates. A
discussion of the approximate MMSE attained by the proposed SPA-DF
structure is included in Appendix II, whereas expressions for the
MMSE of the optimal ordering algorithm are given in Appendix III.
As occurs with S-DF receivers, a disadvantage of the SPA-DF
detector is that it generally does not provide uniform performance
over the user population. In a scenario with tight power control
successive techniques tend to favor the last detected users,
resulting in non-uniform performance. To equalize the performance
of the users an iterative technique with multiple stages can be
used.

\subsection{Iterative Successive Parallel Arbitrated DF Detection}

In \cite{woodward2}, Woodward {\it et al.} presented an iterative
detector with an S-DF in the first stage and P-DF or S-DF
structures, with users being demodulated in reverse order, in the
second stage. The work of \cite{woodward2} was then extended to
account for coded systems and training-based reduced-rank filters
\cite{woodward3}. Here, we focus on the proposed SPA-DF receiver
and the low complexity near-optimal ordering algorithm, and
combine the SPA-DF structure with iterative detection. An
iterative receiver with hard-decision feedback is defined by:
\begin{equation}
{\bf z}^{(m+1)}(i)={\bf W}^{H}(i){\bf r}(i) - {\bf
F}^{H}(i)\hat{\bf b}^{(m)}(i)
\end{equation}
where the filters ${\bf W}$ and ${\bf F}$ can be S-DF or P-DF
structures, and $\hat{\bf b}^{m}(i)$ is the vector of tentative
decisions from the preceding iteration that is described by:
\begin{equation}
\hat{\bf b}^{(1)}(i) = {\rm sgn} \Big(\Re \Big[{\bf W}^{H}(i){\bf
r}(i)\Big] \Big)
\end{equation}
\begin{equation}
\hat{\bf b}^{(m)}(i) = {\rm sgn} \Big(\Re \Big[{\bf
z}^{(m)}(i)\Big] \Big), ~m>1
\end{equation}
where the number of stages $m$ depends on the application. More
stages can be added and the order of the users is reversed from
stage to stage.

To equalize the performance over the user population, we consider
a two-stage structure. The first stage is an SPA-DF scheme with
filters ${\bf W}^{1}$ and ${\bf F}^{1}$. The tentative decisions
are passed to the second stage, which consists of an S-DF, an P-DF
or an SPA-DF detector with filters ${\bf W}^{2}$ and ${\bf
F}^{2}$, that are computed similarly to ${\bf W}^{1}$ and ${\bf
F}^{1}$ but use the decisions of the first stage.  The resulting
iterative receiver system is denoted ISPAS-DF when an S-DF scheme
is deployed in the second stage, whereas for P-DF filters in the
second stage the overall scheme is called ISPAP-DF. The output of
the second stage of the resulting scheme is:
\begin{equation}
z_{j}^{(2)}(i)=[{\bf M}{\bf W}^{2}(i)]_{j}^{H}{\bf r}(i) - [{\bf
M}{\bf F}^{2}(i)]_{j}^{H}\hat{\bf b}^{(2)}(i)
\end{equation}
where $z_{j}$ is the $j$th component of the soft output vector
${\bf z}$, ${\bf M}$ is a square permutation matrix with ones
along the reverse diagonal and zeros elsewhere (similar to ${\bf
M}_{4}$ in (18)), $[. ]_{j}$ denotes the $j$th column of the
argument (a matrix), and $\hat{b}_{j}^{m}(i) = {\rm
sgn}[\Re(z_{j}^{m}(i))]$. The third proposed iterative scheme is
denoted ISPASPA-DF and corresponds to an SPA-DF architecture
employed in both stages. The output of the $l$th branch of its
second stage is:
\begin{equation}
z_{l,j}^{(2)}(i)=[{\bf M}{\bf W}^{2}(i)]_{j}^{H}{\bf r}(i) - [{\bf
M}_{l}{\bf F}^{2}(i)]_{j}^{H}\hat{\bf b}^{(2)}(i)
\end{equation}
where $\hat{b}_{j}^{(2)}(i) = {\rm sgn}\Big[\Re \Big( \arg \min_{1
\leq l \leq L} e_{l,j}^{l}(i) \Big) \Big]$ and $e_{l,j} =
|b_{k}(i) - z_{l,j}(i)|$. Note that the users in the second stage
are demodulated successively and in reverse order relative to the
first branch of the SPA-DF structure (a conventional S-DF). The
role of reversing the cancellation order in successive stages is
to equalize the performance of the users over the population or at
least reduce the performance disparities. Indeed, it provides a
better performance than keeping the same ordering as the last
decoded users in the first stage tend to be favored by the reduced
interference. The rationale is that by using these benefited users
(last decoded ones) as the first ones to be decoded in the second
stage, the resulting performance is improved. Additional stages
can be included, although our studies suggest that the gains in
performance are marginal. Hence, the two-stage scheme is adopted
for the rest of this work.

\section{Successive Parallel Arbitrated DF and Iterative Detection for Coded Systems}

This section is devoted to the description of the proposed SPA-DF
detector and iterative detection schemes for coded systems which
employ convolutional codes with Viterbi and turbo decoding.
Specifically, we present iterative DF detectors based on the
proposed SPA-DF structure which exploits user ordering and combine
the SPA-DF with either the S-DF, the P-DF or another SPA-DF in the
second stage. We show that a reduced number of turbo iterations
can be used with the proposed iterative detector when a
near-optimal user ordering is employed and that savings in
transmitted power are also obtained as compared to previously
reported turbo detectors \cite{alexander1}-\cite{galmal}.

\subsection{Convolutional Codes with Viterbi Decoding}

The structure shown in Fig. 1 can be extended to coded systems by
including a decoder after the selection unit and before the slicer
and an encoder that processes the refined estimates before the
feedback filter ${\bf F}(i)$. For the proposed SPA-DF receiver
structure, users are decoded successively with the aid of the
Viterbi algorithm for each parallel arbitrated branch and then
reencoded with a convolutional encoder and used for interference
cancellation. The motivation for the proposed encoded structure is
that significant gains can be obtained from iterative techniques
with soft cancellation methods and error control coding
\cite{foerster}-\cite{galmal} and from efficient receivers
structures and ordering algorithms such as the novel SPA-DF
detector. The decoding process of the existing S-DF, P-DF and
iterative schemes, namely the ISS-DF and the ISP-DF, are explained
in \cite{woodward2}. The decoding of the proposed iterative
detection schemes that employ the SPA-DF detector (ISPAS-DF,
ISPAP-DF and ISPASPA-DF) resembles the uncoded case, where the
second stage benefits from the enhanced estimates provided by the
first stage that now employs convolutional codes followed by a
Viterbi decoder with branch metrics based on the Hamming distance.
Specifically, the output of the second stage of the resulting
scheme for coded systems is:
\begin{equation}
z_{j}^{(2)}(i)=[{\bf M}{\bf W}^{2}(i)]_{j}^{H}{\bf r}(i) - [{\bf
M}{\bf F}^{2}(i)]_{j}^{H}\hat{\bf b}^{(2)}(i)
\end{equation}
where
\begin{equation}
[{\hat{\bf b}}^{(2)}(i)]_{l} = \left\{ \begin{array}{ll} {\hat
b}_{j}^{(2)} & \textrm{for $l>j$} \\ {\hat b}_{j}^{(1)} &
\textrm{for $l<j$}
\end{array}\right.
\end{equation}
where $[{\hat{\bf b}}^{(2)}(i)]_{l}$ is the $l$th entry of the
decision vector ${\hat{\bf b}}^{(2)}(i)$. Accordingly, the output
of the second stage of the ISPASPA-DF (the SPA-DF architecture is
employed in both stages) is desbribed by:
\begin{equation}
z_{l,j}^{(2)}(i)=[{\bf M}{\bf W}^{2}(i)]_{j}^{H}{\bf r}(i) - [{\bf
M}_{l}{\bf F}^{2}(i)]_{j}^{H}\hat{\bf b}^{(2)}(i)
\end{equation}
where $\hat{b}_{j}^{(2)}(i) = {\rm sgn}\Big[\Re \Big( \arg \min_{1
\leq l \leq L} e_{l,j}(i) \Big) \Big]$ and
\begin{equation}
e_{l,j}(i) = \left\{ \begin{array}{ll} |b_{j}^{(2)}(i) -
z_{l,j}(i)| & \textrm{for $l>j$} \\ |b_{j}^{(1)}(i) - z_{l,j}(i)|
& \textrm{for $l<j$}
\end{array}\right.
\end{equation}

\subsection{Iterative Turbo Receiver and Decoding}

\begin{figure*}[t]
\begin{center}
\def\epsfsize#1#2{2.2\columnwidth}
\epsfbox{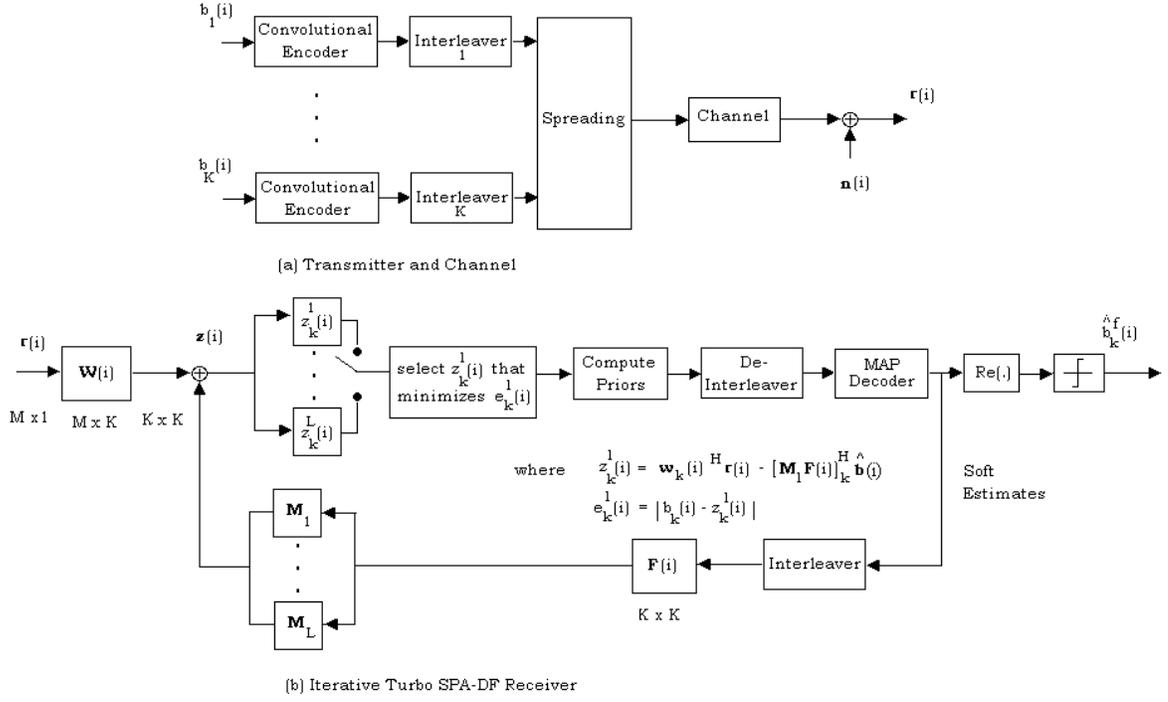} \caption{Block diagram of the proposed system
with the SPA-DF detector and turbo decoding.}
\end{center}
\end{figure*}

A CDMA system with convolutional codes being used at the
transmitter and the proposed iterative SPA-DF receiver with turbo
decoding is illustrated in Fig. 2. The proposed iterative (turbo)
receiver structure consists of the following stages: a
soft-input-soft-output (SISO) SPA-DF detector and a maximum
\textit{a posteriori} (MAP) decoder. These stages are separated by
interleavers and deinterleavers. Specifically, soft outputs from
the SPA-DF are used to estimate likelihoods which are interleaved
and input to the MAP decoder for the convolutional code. The MAP
decoder computes \textit{a posteriori} probabilities (APPs) for
each user's encoded symbols, which are used to generate soft
estimates. These soft estimates are subsequently used to update
the SPA-DF filters, de-interleaved and fed back through the
feedback filter. This process is then iterated.

The proposed SPA-DF detector yields the \textit{a posteriori}
log-likelihood ratio (LLR) of a transmitted symbol ($+1$ or $-1$)
for every code bit of each user as given by
\begin{equation}
\Lambda_1[b_k(i)] = {\rm log} \frac{P[b_k(i)=+1|{\bf
r}(i)]}{P[b_k(i)=-1| {\bf r}(i)]}, ~~~k=1, \ldots, K.
\end{equation}
Using Bayes' rule, the above equation can be written as
\begin{equation}
\begin{split}
\Lambda_1[b_k(i)] & = {\rm log} \frac{P[{\bf r}(i)|b_k(i)=+1]}{P[
{\bf r}(i)|b_k(i)=-1]} + {\rm log}
\frac{P[b_k(i)=+1]}{P[b_k(i)=-1]} \\ & = \lambda_1[b_k(i)] +
\lambda_2^p[b_k(i)]
\end{split}
\end{equation}
where $\lambda_2^p[b_k(i)] = {\rm log}
\frac{P[b_k(i)=+1]}{P[b_k(i)=-1]}$ represents the \textit{a
priori} LLR of the code bit $b_k(i)$, which is computed by the MAP
decoder of the $k$th user in the previous iteration, interleaved
and then fed back to the SPA-DF detector. Note that the
superscript $^p$ denotes the quantity obtained in the previous
iteration. Assuming equally likely bits, for the first iteration
we have $\lambda_2^p[b_k(i)] =0$ for all users. The first term in
(33), i.e. $\lambda_1[b_k(i)] = {\rm log} \frac{P[{\bf
r}(i)|b_k(i)=+1]}{P[ {\bf r}(i)|b_k(i)=-1]}$, represents the
\textit{extrinsic} information yielded by the SISO SPA-DF detector
based on the received data ${\bf r}(i)$, the prior information
about the code bits of all other users $\lambda_2^p[b_l(i)], l
\neq k$ and the prior information about the code bits of the $k$th
user other than the $i$th bit. The extrinsic information
$\lambda_1[b_k(i)]$ provided by the MAP decoder is then
de-interleaved and fed back into the MAP decoder of the $k$th user
as the \textit{a priori} information in the next iteration.

Based on the prior information $\lambda_1^p[b_k(i)]$ and the
trellis structure of the code, the $k$th user's MAP decoder
computes the \textit{a posteriori} LLR of each code bit as
described by
\begin{equation}
\begin{split}
\Lambda_2[b_k(i)] & = {\rm log} \frac{P[b_k(i)=+1|
\lambda_1^p[b_k(i); {\rm decoding}]}{P[b_k(i)=-1|
\lambda_1^p[b_k(i); {\rm decoding}]}
\\ & = \lambda_2[b_k(i)] + \lambda_1^p[b_k(i)] , ~~~k=1, \ldots, K.
\end{split}
\end{equation}
From the above equality, it is seen that the output of the MAP
decoder is the sum of the prior information $\lambda_1^p[b_k(i)]$
and the extrinsic information $\lambda_2[b_k(i)]$ yielded by the
MAP decoder. This extrinsic information is the information about
the code bit $b_k(i)$  obtained from the prior information about
the other code bits $\lambda_1^p[b_k(j)], ~ j \neq i$ \cite{wang}.
The MAP decoder also computes the \textit{a posteriori} LLR of
every information bit, which is used to make a decision on the
decoded bit at the last iteration. After interleaving, the
extrinsic information yielded by the $K$ MAP decoders
$\lambda_2[b_k(i)], ~k=1, \ldots,K$ is fed back to the SPA-DF
detector, as the prior information about the code bits of all
users in the subsequent iteration. At the first iteration, the
extrinsic information $\lambda_1[b_k(i)]$ and $\lambda_2[b_k(i)]$
are statistically independent and as the iterations are computed
they become more correlated and the improvement due to each
iteration is gradually reduced.

For the purpose of MAP decoding, we assume that the interference
plus noise at the output of the subtractor in Fig. 2 (b), which
corresponds to ${\bf z}(i)$, is Gaussian. This assumption is
reasonable when there are many active users, has been used in
previous works \cite{woodward3},\cite{wang}-\cite{galmal} and
provides an efficient and accurate way of computing the extrinsic
information. Thus, for the $k$th user and $m$th iteration the soft
output of the SPA-DF detector is written as
\begin{equation}
z_k^{(m)}(i) = V_k^{(m)} b_k(i) + \xi_k^{(m)}(i)
\end{equation}
where $V_k^{(m)}(i)$ is a scalar variable equivalent to the $k$th
user's amplitude and $\xi_k^{(m)}(i)$ is a Gaussian random
variable with variance $\sigma^2_{\xi_k^{(m)}}$. Since we have
\begin{equation}
V_k^{(m)}(i) = E\big[ b_k^*(i) z_k^{(m)}(i) \big]
\end{equation}
and
\begin{equation}
\sigma^2_{\xi_k^{(m)}}(i)  = E\big[ | z_k^{(m)}(i) -  V_k^{(m)}(i)
b_k(i)|^2 \big]
\end{equation}
the designer can obtain the estimates ${\hat V}_k^{(m)}(i)$ and
${\hat \sigma}^2_{\xi_k^{(m)}}(i)$ via the corresponding sample
averages over the packet transmission. These estimates are used to
compute the detector \textit{a posteriori} probabilities $P[b_k(i)
= \pm 1 | z_k^{(m)}(i)]$ which are de-interleaved and input to the
MAP decoder for the convolutional code. In what follows, we assume
that the MAP decoder generates APPs $P[b_k(i) = \pm 1]$, which are
used to compute the input to the feedback filter ${\bf f}_k(i)$.
From (35) the extrinsic information delivered by the soft output
SPA-DF is given by
\begin{equation}
\begin{split}
\lambda_1[b_k(i)] & = {\rm log} \frac{P[z_k^{(m)}(i)|b_k(i)=+1]}{
P[z_k^{(m)}(i)|b_k(i)=-1] } = -\frac{(z_k^{(m)}(i) -
V_k^{(m)})^2}{2 \sigma^2_{\xi_k^{(m)}}(i)} \\ & \quad +
\frac{(z_k^{(m)}(i) + V_k^{(m)})^2}{2 \sigma^2_{\xi_k^{(m)}}(i)} =
\frac{2V_k^{(m)} z_k^{(m)}(i)}{ \sigma^2_{\xi_k^{(m)}}(i) }
\end{split}
\end{equation}
The SPA-DF turbo detector chooses the best estimate of the $L$
candidates for the $m$th turbo decoding iteration as:
\begin{equation}
l_{best,k}^{(m)}(i) = \arg \min_{1 \leq l \leq L} e_{k}^{l}(i)
\end{equation}
where the best estimate is the value $z_{k}^{l}(i)$ which
minimizes $e_{k}^{l}(i)=|b_{k}(i) - z_{k}^{l}(i)|$.

\subsection{Extensions}

Here, we briefly comment on how the proposed receiver structures
can be extended to take into account asynchronous systems, dynamic
scenarios, other types of communications systems and multiple
access techniques.

For asynchronous systems with large relative delays amongst the
users, the observation window of each user should be expanded in
order to consider an increased number of samples derived from the
offsets amongst users. Alternatively for small relative delays
amongst users, the designer can resort to chip oversampling to
compensate for the random timing offsets. These remedies imply in
augmented filter lengths and consequently increased computational
complexity. To alleviate for the increase in filter length and the
increased amount of training, the designer can resort to
reduced-rank estimation techniques such as the Multistage Wiener
Filter, as in \cite{woodward2}, or to a new very promising
technique that employs interpolated FIR filters \cite{delamaresp}.

An extension with low complexity turbo schemes such as the one in
\cite{vogelbruch} are also possible with the structures presented
in this paper. For dynamic channels that are subject to fading,
the designer can rely on adaptive signal processing techniques and
make the proposed detector structures adaptive in order to track
the variations of the channel and the interference. This includes
some modifications for CDMA systems with long codes, which require
a different approach for estimating the covariance observation
matrix ${\bf R}$ due to the loss of the cyclostationarity.

Finally, we also remark that the proposed detection schemes can be
deployed for narrow-band systems with multiple transmitter and
receiver antennas, exploiting the capacity improvements of spatial
multiplexing.

\section{Simulations}

In this section, we evaluate the performance of the iterative
arbitrated DF structures introduced in Section IV and compare them
with other existing structures. Due to the extreme difficulty of
theoretically analyzing such scheme, we adopt a simulation
approach and conduct several experiments in order to verify the
effectiveness of the proposed techniques. In particular, we have
carried out experiments to assess the bit error rate (BER)
performance of the DF receivers for different loads, channel
profiles, and signal to noise ratios ($E_{b}/N_{0}$). The DS-CDMA
system employs random generated spreading sequences of length
$N=16$, $N=32$ and $N=64$, has perfect power control and use
statistically independent random channels with $L_{p}=3$, whose
coefficients $h_{k,l}$ are taken, for each run, from uniform
random variables between $-1$ and $1$, and which are normalized so
that $\sum_{l=1}^{L_{p}} h_{k,l}^{2}=1$. It should be remarked
that the existence of multipath creates an error floor for the
multiuser receivers, making it more difficult the interference
suppression of associated users. Note also that given the
performance of current power control algorithms, ideal power
control is not far from a realistic situation. The matrices used
in (14) and (15) are estimated by ${\hat{\bf R}}(i) =
\frac{1}{i}\sum_{l=1}^{i} {\bf r}(l){\bf r}^{H}(l)$ and ${\hat{\bf
B}}(i) = \frac{1}{i} \sum_{l=1}^{i} {\bf r}(l){\hat {\bf
b}}^{H}(l)$. For coded systems, we employ a convolutional code
with rate $R=3/4$ and constraint length $6$ which can be found in
\cite{proakis}. In particular, for turbo decoding plots we used
S-random interleavers with block size equal to $256$. In the
following experiments, averaged over $200$ runs for uncoded
systems, over $2000$ for encoded systems with Viterbi decoding and
over $20000$ for turbo decoded schemes, it is indicated the
receiver structure (linear or decision feedback (DF)). Amongst the
different DF structures, we consider:

\begin{itemize}

\item{S-DF: the successive DF detector of \cite{duel,varanasi2}.}

\item{P-DF: the parallel DF detector of \cite{woodward1,woodward2}.}

\item{ISS-DF: the iterative system of Woodward {\it et al.} \cite{woodward2}
with S-DF in the first and second stages.}
\item{ISP-DF: the iterative system of Woodward {\it et al.} \cite{woodward2}
with S-DF in the first stage and P-DF in the second stage.}
\item{SPA-DF: the proposed successive parallel arbitrated receiver.}
\item{ISPAS-DF: the proposed iterative detector with the novel SPA-DF
in the first stage and the S-DF in the second stage.}
\item{ISPAP-DF: the proposed iterative receiver with the SPA-DF in the
first stage and the P-DF in the second stage.}
\item{ISPASPA-DF: the proposed iterative receiver with the SPA-DF in the
first and second stages.}

\end{itemize}

\begin{figure}[!htb]
\begin{center}
\def\epsfsize#1#2{1\columnwidth}
\epsfbox{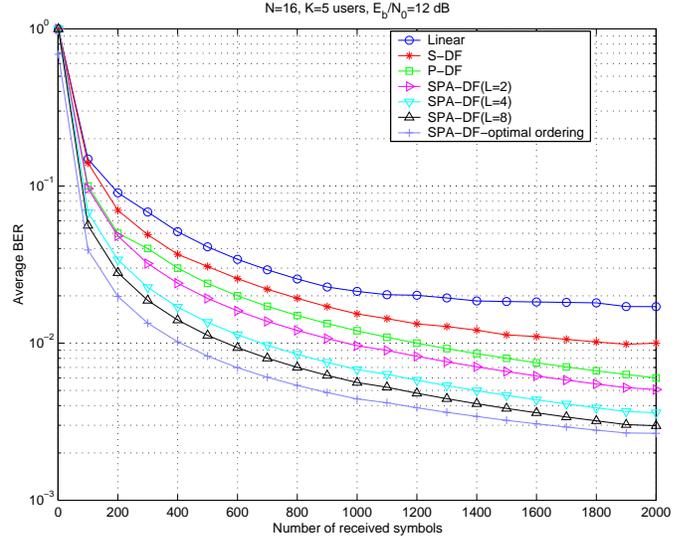} \caption{BER performance versus number of
symbols.}
\end{center}
\end{figure}

Let us first consider the proposed SPA-DF, evaluate the number of
arbitrated branches that should be used in the ordering algorithm
and account for the impact of additional branches upon
performance. In addition to this, we carry out a comparison of the
proposed low complexity user ordering algorithm against the
optimal ordering approach, briefly described in Section IV. A,
that tests $K!$ possible branches and selects the most likely
estimate. We designed the novel DF receivers with $L=2,4,8$
parallel branches and compared their BER performance versus number
of symbols with the existing S-DF and P-DF structures, as depicted
in Fig. 3. The results show that the proposed low complexity
ordering algorithm achieves a performance close to the optimal
ordering, whilst keeping the complexity reasonably low for
practical utilization. Furthermore, the performance of the new
SPA-DF scheme with $L=2,4,8$ outperforms the S-DF and the P-DF
detector. It can be noted from the curves that the performance of
the new SPA-DF improves as the number of parallel branches
increase. In this regard, we also notice that the gains of
performance obtained through additional branches decrease as $L$
is increased, resulting in marginal improvements for more than
$L=4$ branches. For this reason, we adopt $L=4$ for the remaining
experiments because it presents a very attractive trade-off
between performance and complexity.

A performance comparison in terms of BER of the proposed DF
structures, namely SPA-DF, ISPAP-DF, ISPAS-DF and ISPASPA-DF with
existing iterative and conventional DF and linear detectors is
illustrated in Figs. 4 to 5, for uncoded systems and in Fig. 6,
for convolutionally coded systems. In particular, we show BER
performance curves versus $E_{b}/N_{0}$ and number of users (K)
for the analyzed receivers. The results for a system with $N=32$,
depicted in Fig. 4 indicate that the best performance is achieved
with the novel ISPASPA-DF (the SPA-DF is employed in two cascaded
stages), followed by the new ISPAP-DF, the existing ISP-DF
\cite{woodward2}, the ISPAS-DF, the SPA-DF, the P-DF, the ISS-DF,
the S-DF and the linear detector. Specifically, the ISPASPA-DF
detector can save up to $1.5$ dB and support up to $4$ more users
in comparison with the ISP-DF (which is the best existing scheme)
for the same BER performance. The ISPAP-DF scheme can save up to
$1$ dB and support up to $2$ more users in comparison with the
ISP-DF for the same BER performance. Moreover, the performance
advantages of the ISPASPA-DF and ISPAP-DF systems are
substantially superior to the other existing approaches.

\begin{figure}[!htb]
\begin{center}
\def\epsfsize#1#2{1\columnwidth}
\epsfbox{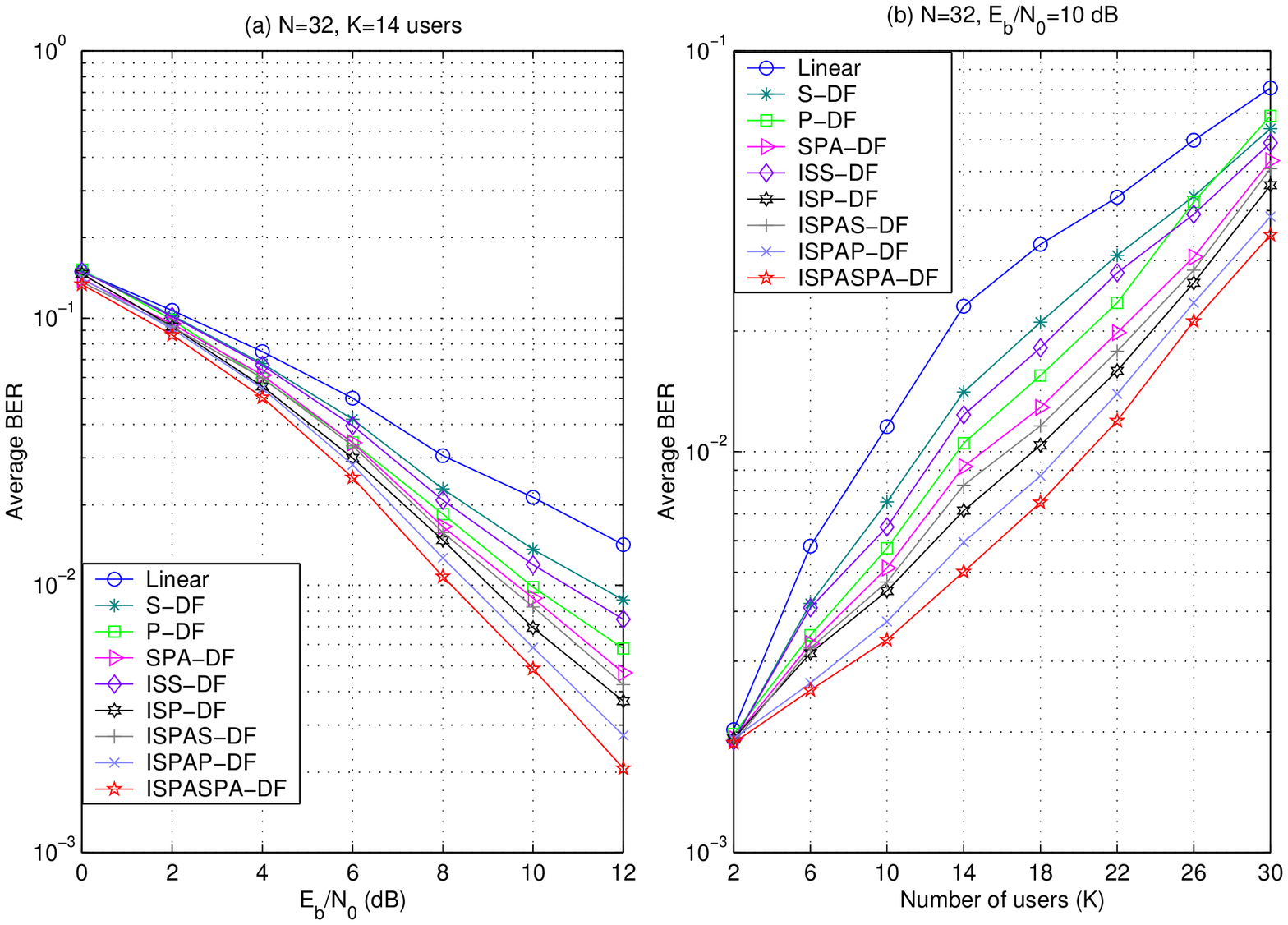} \caption{BER performance versus (a)
$E_{b}/N_{0}$ and (b) number of users (K).}
\end{center}
\end{figure}

The results for a larger system with $N=64$, illustrated in Fig.
5, corroborate the curves obtained for the smaller system in Fig.
4. In particular, the same BER performance hierarchy is observed
for the detection schemes (except for the ISPAS-DF, that now
outperforms the ISP-DF) and we notice some additional gains in
performance for the proposed schemes over the existing techniques.
Specifically, the ISPASPA-DF detector can save up to $1.8$ dB and
support up to $10$ additional users in comparison with the ISP-DF
for the same BER performance. The ISPAP-DF scheme can save up to
$1.4$ dB and support up to $8$ more users in comparison with the
ISP-DF for the same BER performance. Moreover, the performance
advantages of the ISPASPA-DF and ISPAP-DF systems are even more
pronounced over the other analyzed schemes for larger systems.

\begin{figure}[!htb]
\begin{center}
\def\epsfsize#1#2{1\columnwidth}
\epsfbox{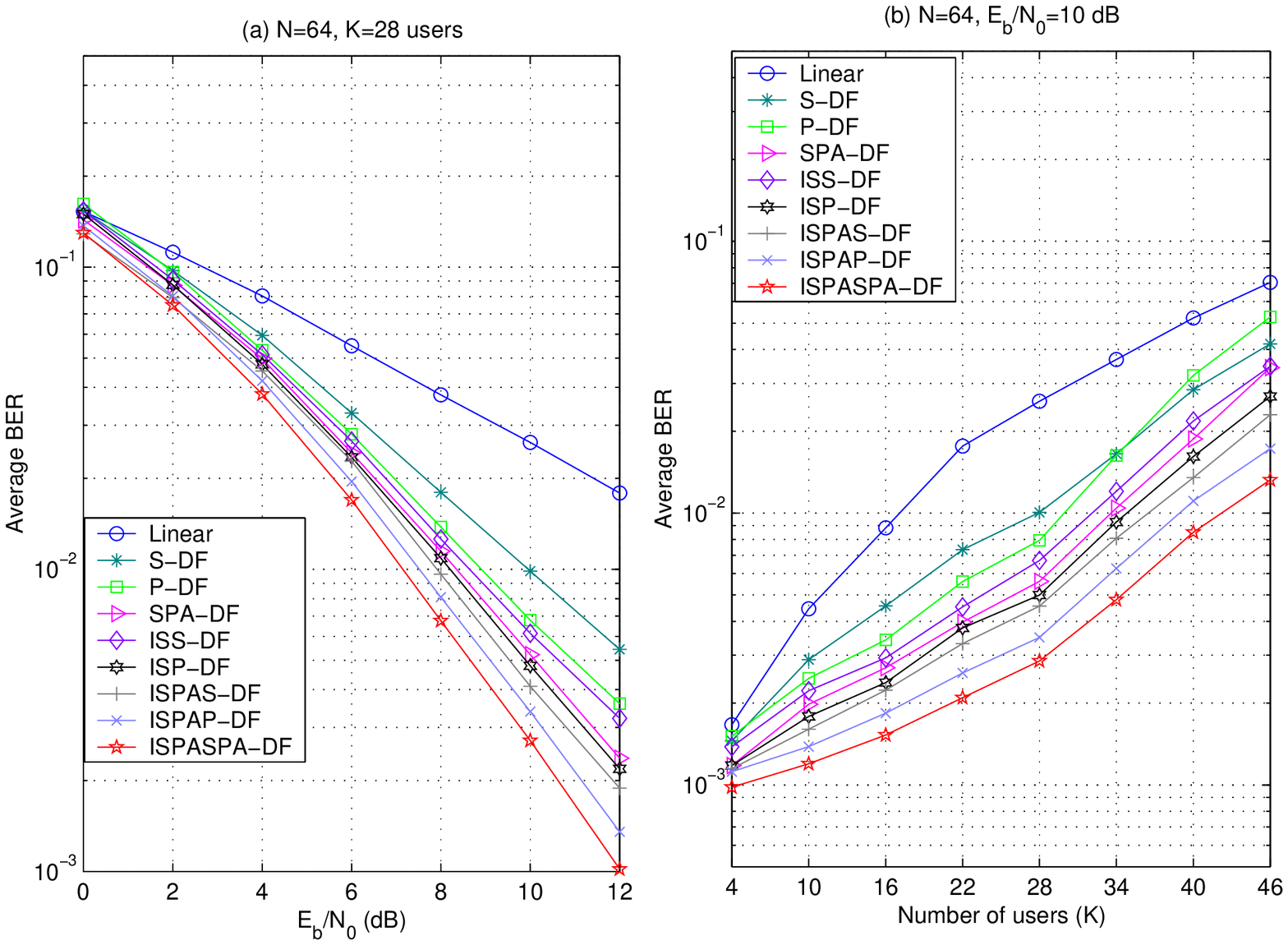} \caption{BER performance versus (a)
$E_{b}/N_{0}$ and (b) number of users (K).}
\end{center}
\end{figure}

The BER performance of the analyzed detection schemes was then
examined for convolutionally encoded systems with Viterbi
decoding, $N=32$ and rate $R=3/4$, as depicted in Fig. 6. The
results corroborate those obtained for uncoded systems in Figs. 4
and 5, and indicate that the proposed ISPASPA-DF and ISPAP-DF
detection schemes significantly outperform the remaining receiver
structures. In particular, the ISPASPA-DF detector can support up
to $8$ additional users in comparison with the ISP-DF for the same
BER performance, whereas the ISPAP-DF scheme can accomodate up to
$6$ more users in comparison with the ISP-DF for the same BER
performance. It is worth noting that the linear and P-DF detectors
experience performance losses for coded systems, relative to the
other structures, as verified in \cite{woodward2} and which is a
result of the loss in spreading gain that increases the
interference power at the output of the MMSE receiver.

\begin{figure}[!htb]
\begin{center}
\def\epsfsize#1#2{1\columnwidth}
\epsfbox{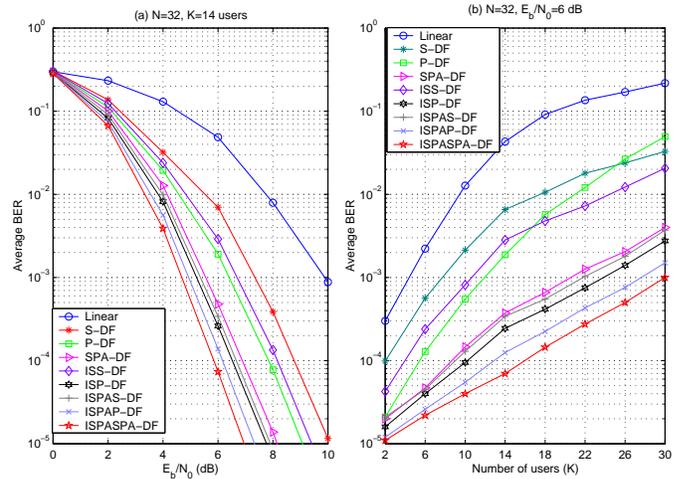} \caption{BER performance of a convolutionally
coded system with $R=3/4$ versus (a) $E_{b}/N_{0}$ and (b) number
of users (K).}
\end{center}
\end{figure}

The BER performance of the analyzed detection schemes was also
investigated for convolutionally encoded systems with turbo
decoding. In our studies with turbo receivers, we tested several
code rates and found that $R=1/2$ was unable to attain good
performance for highly loaded systems, whereas $R=3/4$ was
powerful enough to obtain good performance even in fully loaded
systems. For this reason, we adopted the rate $R=3/4$ for the
remaining experiments with iterative decoders and considered a
system with $N=32$, as depicted in Fig. 7. The results corroborate
those obtained for uncoded and encoded systems with Viterbi
decoding in Figs. 5 and 6, and indicate that the proposed
ISPASPA-DF and ISPAP-DF detection schemes significantly outperform
the remaining receiver structures. In particular, the ISPASPA-DF
detector can approach the single user bound with only $4$
iterations and offer a significant advantage over the existing
detectors. In comparison with existing iterative DF detectors, the
ISPASPA-DF can save up to $0.5$ dB for the same BER performance,
whereas it can accommodate a fully loaded system with only $4$
iterations and operating with only $4$ dB with negligible
performance degradation as the load is increased.

\begin{figure}[!htb]
\begin{center}
\def\epsfsize#1#2{1\columnwidth}
\epsfbox{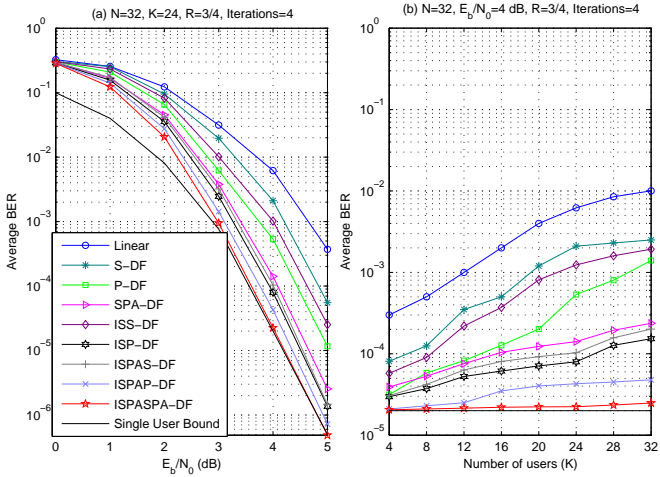} \caption{BER performance of a turbo decoded
system with $R=3/4$ versus (a) $E_{b}/N_{0}$ and (b) number of
users (K).}
\end{center}
\end{figure}

In Fig. 8 it is illustrated the average BER performance of the
detectors versus the number of iterations of the turbo decoder.
The plots show that the proposed ISPASPA-DF and the ISPAP-DF
detectors achieve the single user bound with only $4$ and $7$
iterations, respectively, whereas the remaining detectors require
more iterations to achieve this performance. This is an important
feature of the proposed detectors as they can save considerable
computational resources by operating with a lower number of turbo
iterations.

\begin{figure}[!htb]
\begin{center}
\def\epsfsize#1#2{1\columnwidth}
\epsfbox{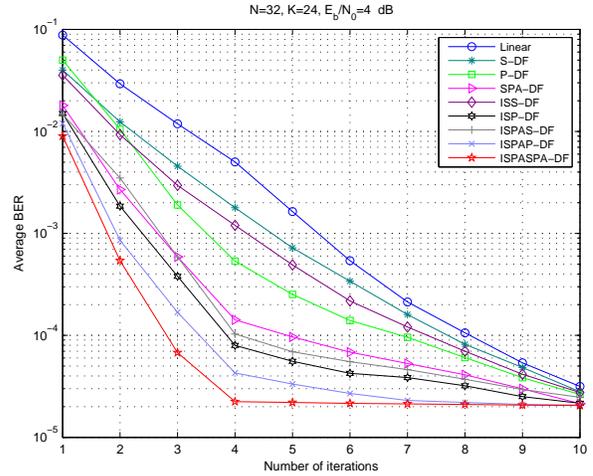} \caption{BER performance of a turbo decoded
system with $R=3/4$ versus number of iterations.}
\end{center}
\end{figure}

The last scenario, shown in Figs. 9, considers the individual BER
performance of the users for both uncoded and convolutionally
encoded systems with Viterbi decoding. From the curves, we observe
that a disadvantage of S-DF relative to P-DF is that it does not
provide uniform performance over the user population. We also
notice that for the S-DF receivers, user $1$ achieves the same
performance of their linear receivers counterparts, and as the
successive cancellation is performed users with higher indices
benefit from the interference cancellation. The same non-uniform
performance is verified for the proposed SPA-DF, the existing
ISS-DF and the novel ISPAS-DF and ISPASPA-DF. Conversely, the new
ISPAP-DF, the existing P-DF and the existing ISP-DF provide
uniform performance over the users which is an important goal for
the uplink of DS-CDMA systems. In particular, the novel ISPAP-DF
detector achieves the best uniform performance of the analyzed
structures and is superior to the ISP-DF and to the P-DF, that
suffers from error propagation. For coded systems, we notice that
the performance of the proposed ISPASPA-DF and ISPAS-DF, and the
existing ISS-DF and S-DF becomes very attractive for the users
with indices greater than $5$ (where the SIC-based schemes
outperform the ISPAP-DF, the ISP-DF and the P-DF). This suggests
the deployment of these structures for systems that rely on
differentiated services, where the quality of service (QoS) can be
made different for different groups of users. In this context and
as an example, users with the first indices and poorer performance
should be allocated to voice services, while the users with better
performance should be designated to data transmission services
that require improved QoS.

\begin{figure}[!htb]
\begin{center}
\def\epsfsize#1#2{1\columnwidth}
\epsfbox{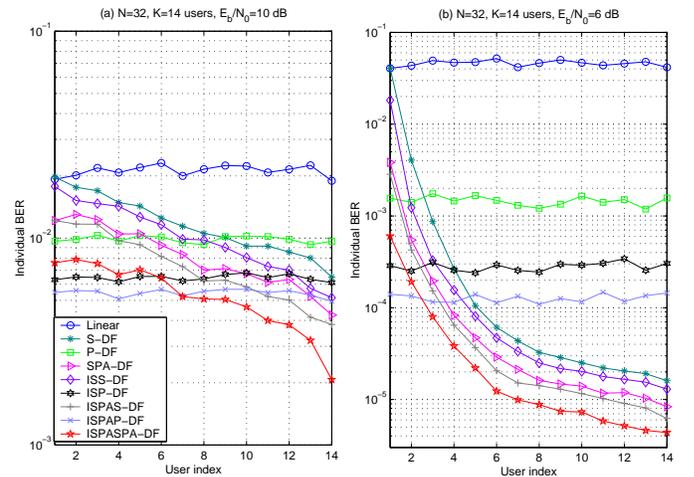} \caption{BER performance versus user index for
(a) an uncoded system (b) a convolutionally coded system with rate
$R=3/4$.}
\end{center}
\end{figure}

\section{Conclusions}

A novel SPA-DF structure and a low complexity near-optimal
ordering algorithm were presented and combined with iterative
techniques for use with cascaded DF stages for mitigating the
deleterious effects of error propagation. The proposed SPA-DF and
iterative receivers for DS-CDMA systems were investigated in an
uplink scenario and compared to existing schemes in the
literature. The results for both uncoded and convolutionally
encoded systems using Viterbi and turbo decoding show that the new
detection schemes can offer considerable gains as compared to
existing DF and linear receivers, support systems with higher
loads and mitigate the phenomenon of error propagation.

\begin{appendix}

\section{Relationships between the MMSE with perfect and imperfect
feedback for DF detectors }

In this Appendix, we provide some relationships between the MMSE
attained by a decision feedback structure with perfect and
imperfect feedback. Let us consider an alternative expression for
the cost function in (4) for user $k$:
\begin{equation}
\begin{split}
J_{MSE} & = \sigma^2_{b} - {\bf w}_{k}^{H}{\bf p}_{k} - {\bf
p}_{k}^{H}{\bf w}_{k} + {\bf w}_{k}^{H}{\bf R}{\bf w}_{k} + {\bf
f}_{k}^{H}{\bf f}_{k} - {\bf w}_{k}^{H}{\bf B}{\bf f}_{k} \\ &
\quad - {\bf f}^{H}{\bf B}^{H}{\bf w}_{k}
\end{split}
\end{equation}
Consider the expression for the feedforward filter ${\bf w}_{k}=
{\bf R}^{-1}({\bf p}_{k} + {\bf B}{\bf f}_{k})$ obtained in (16)
and the expression for the feedback filter ${\bf f}_{k}= {\bf
Q}^{-1} {\bf B}^{H}{\bf w}_{k}$ with ${\bf Q} = E[\hat{\bf
b}\hat{\bf b}^{H}]$ in (17).  By substituting the optimal MMSE
expressions obtained in (17) into (16) for the filters we obtain
an alternative expression for the feedback filter ${\bf f}_{k}$:
\begin{equation}
{\bf f}_{k}= {\bf D}^{-1}{\bf Q}^{-1} {\bf B}^{H}{\bf R}^{-1}{\bf
p}_{k}
\end{equation}
where ${\bf D} = ({\bf I} - {\bf Q}^{-1}{\bf B}^{H}{\bf
R}^{-1}{\bf B})$ and the above expression only depends on ${\bf
Q}$, ${\bf B}$, ${\bf R}$ and ${\bf p}_{k}$. By inserting the
expression ${\bf w}_{k}= {\bf R}^{-1}({\bf p}_{k} + {\bf B}{\bf
f}_{k})$ and (41) into (40), we have for user $k$:
\begin{equation}
\begin{split}
J_{MMSE}  & = \sigma^2_{b} - {\bf p}_{k}^{H}{\bf R}^{-1}{\bf
p}_{k} - {\bf f}_{k}^{H}{\bf B}^{H}{\bf R}^{-1}{\bf p}_{k} - {\bf
p}_{k}^{H}{\bf R}^{-1}{\bf B}{\bf f}_{k}  \\ & \quad -{\bf
f}_{k}^{H}{\bf B}^{H}{\bf R}^{-1}{\bf B}{\bf f}_{k} + {\bf f}_{k}^{H}{\bf f}_{k} \\
& = \sigma^2_{b} - {\bf p}_{k}^{H}{\bf R}^{-1}{\bf p}_{k} - {\bf
p}_{k}^{H}{\bf R}^{-1}{\bf B}{\bf Q}^{-1}{\bf D}^{-1}{\bf
B}^{H}{\bf R}^{-1}{\bf p}_{k} \\ & \quad - {\bf p}_{k}^{H}{\bf
R}^{-1}{\bf B}{\bf Q}^{-1}{\bf D}^{-1}{\bf Q}^{-1}{\bf B}^{H}{\bf R}^{-1}{\bf p}_{k} \\
& \quad  - {\bf p}_{k}^{H}{\bf R}^{-1}{\bf B}{\bf Q}^{-1}{\bf
D}^{-1}{\bf B}^{H}{\bf R}^{-1}{\bf B}{\bf D}^{-1}{\bf Q}^{-1}{\bf
B}^{H}{\bf R}^{-1}{\bf p}_{k} \\ & \quad + {\bf p}_{k}^{H}{\bf
R}^{-1}{\bf B}{\bf Q}^{-1}{\bf D}^{-1}{\bf Q}^{-1}{\bf B}^{H}{\bf
R}^{-1}{\bf p}_{k} \\ & = \sigma^2_{b} - {\bf p}_{k}^{H}{\bf
R}^{-1}{\bf p}_{k} - {\bf p}_{k}^{H}{\bf R}^{-1}{\bf B}{\bf
Q}^{-1}{\bf D}^{-1}{\bf B}^{H}{\bf R}^{-1}{\bf p}_{k} \\ & \quad -
{\bf p}_{k}^{H}{\bf
R}^{-1}{\bf B}{\bf Q}^{-1}{\bf D}^{-1}{\bf Q}^{-1}{\bf B}^{H}{\bf R}^{-1}{\bf p}_{k} \\
& \quad + {\bf p}_{k}^{H}{\bf R}^{-1}{\bf B}{\bf Q}^{-1}{\bf
D}^{-1}({\bf I}-{\bf B}^{H}{\bf R}^{-1}{\bf B}){\bf D}^{-1}{\bf
Q}^{-1}{\bf B}^{H}{\bf R}^{-1}{\bf p}_{k}
\end{split}
\end{equation}
At this point, it is convenient to adopt the judicious
approximation ${\bf Q}=E[\hat{\bf b}\hat{\bf b}^{H}] \approx {\bf
I}$, which is justified for moderate to low BER values. By using
this approximation we have ${\bf f}_{k} \approx {\bf D}^{-1}{\bf
B}^{H}{\bf R}^{-1}{\bf p}_{k}$, where ${\bf D} \approx ({\bf I} -
{\bf B}^{H}{\bf R}^{-1}{\bf B})$, and the MMSE expression for user
$k$ is approximated by:
\begin{equation}
\begin{split}
J_{MMSE}   & \approx \sigma^2_{b} - {\bf p}_{k}^{H}{\bf
R}^{-1}{\bf p}_{k} - {\bf p}_{k}^{H}{\bf R}^{-1}{\bf B}{\bf
D}^{-1}{\bf B}^{H}{\bf R}^{-1}{\bf p}_{k} \\
& \quad  - {\bf p}_{k}^{H}{\bf
R}^{-1}{\bf B}{\bf D}^{-1}{\bf B}^{H}{\bf R}^{-1}{\bf p}_{k} \\
& \quad + {\bf p}_{k}^{H}{\bf R}^{-1}{\bf B}{\bf D}^{-1}({\bf
I}-{\bf B}^{H}{\bf R}^{-1}{\bf B}){\bf D}^{-1}{\bf B}^{H}{\bf
R}^{-1}{\bf p}_{k} \\ & \approx \sigma^2_{b} - {\bf p}_{k}^{H}{\bf
R}^{-1}{\bf p}_{k} - {\bf p}_{k}^{H}{\bf R}^{-1}{\bf B}{\bf
D}^{-1}{\bf B}^{H}{\bf R}^{-1}{\bf p}_{k} \\
& \quad  - {\bf p}_{k}^{H}{\bf
R}^{-1}{\bf B}{\bf D}^{-1}{\bf B}^{H}{\bf R}^{-1}{\bf p}_{k} \\
& \quad + {\bf p}_{k}^{H}{\bf R}^{-1}{\bf B}{\bf D}^{-1}{\bf
B}^{H}{\bf R}^{-1}{\bf p}_{k} \\ & \approx \sigma^2_{b} - {\bf
p}_{k}^{H}{\bf R}^{-1}{\bf p}_{k} - {\bf p}_{k}^{H}{\bf
R}^{-1}{\bf B}{\bf D}^{-1}{\bf B}^{H}{\bf R}^{-1}{\bf p}_{k} \\ &
\approx \sigma^2_{b} - {\bf p}_{k}^{H}{\bf R}^{-1}{\bf p}_{k} \\
& \quad  - {\bf p}_{k}^{H}{\bf R}^{-1}{\bf B}({\bf I} - {\bf
B}^{H}{\bf R}^{-1}{\bf B})^{-1}{\bf B}^{H}{\bf R}^{-1}{\bf p}_{k}
\end{split}
\end{equation}
The approximate expression obtained in (43) represents the MMSE
attained by a general decision feedback structure that has
imperfect feedback. The equation in (43) is a function of ${\bf
B}$, ${\bf R}$ and ${\bf p}_{k}$, and is still dependent on the
decisions. Let us now assume perfect feedback (${\bf b} = \hat{\bf
b}$) and look at the filter expressions. Since ${\bf w}_{k} = {\bf
R}^{-1}({\bf p}_{k} + {\bf B} {\bf f}_{k})$ and ${\bf f}_{k} =
{\bf B}^{H}{\bf w}_{k} = {\bf p}^{H}{\bf R}^{-1}({\bf p}_{k} +
{\bf B} {\bf f}_{k})$
\begin{equation}
\begin{split}
J_{MMSE}   & \approx \sigma^2_{b} - {\bf p}_{k}^{H}{\bf
R}^{-1}{\bf p}_{k} - {\bf p}_{k}^{H}{\bf R}^{-1}{\bf B}{\bf
D}^{-1}{\bf B}^{H}{\bf R}^{-1}{\bf p}_{k} \\ & \approx
\sigma^2_{b} - {\bf p}_{k}^{H}{\bf R}^{-1}{\bf p}_{k} - {\bf
p}_{k}^{H}{\bf R}^{-1}{\bf B}{\bf f}_{k}\\ & \approx \sigma^2_{b}
- {\bf p}_{k}^{H}{\bf R}^{-1}({\bf p}_{k} + {\bf B}{\bf f}_{k})
\\ & \approx \sigma^2_{b} - {\bf p}_{k}^{H}{\bf R}^{-1}{\bf R}{\bf
w}_{k} = \sigma^2_{b} - {\bf p}_{k}^{H}{\bf w}_{k}
\end{split}
\end{equation}
The approximate expression obtained in (43) has been significantly
simplified due to the assumption of perfect feedback and indicates
that the MMSE for user $k$ is a function of ${\bf w}_{k}$. If we
consider a decision feedback structure such as successive
cancellation (S-DF), use the expression for the feedforward filter
${\bf w}_{k} = {\bf R}^{-1}_{U}{\bf p}_{k}$, the MMSE for user $k$
is approximately given by:
\begin{equation}
J_{MMSE}  \approx  \sigma^2_{b} - {\bf p}_{k}^{H}{\bf
R}^{-1}_{U}{\bf p}_{k}
\end{equation}
where the above result means that the MMSE attained by user $k$ is
proportional to the number of undetected users expressed by the
covariance matrix ${\bf R}_{U}$. If we consider a decision
feedback structure such as parallel cancellation (P-DF), use the
expression for the feedforward filter ${\bf w}_{k} = {\bf
R}^{-1}_{U}{\bf p}_{k} = \frac{{\bf p}_{k}}{|A_{k}|^{2} +
\sigma^{2}}$, the MMSE for user $k$ is approximately given by:
\begin{equation}
J_{MMSE}  \approx  \sigma^2_{b} - {\bf p}_{k}^{H}({\bf p}_{k}{\bf
p}_{k}^{H} + \sigma^{2}{\bf I})^{-1}{\bf p}_{k}
\end{equation}
Note that the above result corresponds to the single-user bound
because we assume that all users (with perfect decision) had been
fed back, as in P-DF.

\section{On the MMSE of the Proposed SPA-DF detectors }

For imperfect feedback, the P-DF is known to be susceptible to
error propagation, while the S-DF is more effective in combating
these deleterious effects. The proposed SPA-DF employs several
versions of S-DF in parallel and chooses the best estimate amongst
these parallel branches, resulting in improved performance over
the S-DF, as verified in our studies. Here, we mathematically
discuss the MMSE of the SPA-DF, under the assumption of perfect
feedback. If we consider the SPA-DF with $L$ branches, we have $L$
different groups of undetected users, namely,
$U_{1},~U_{2},~\ldots,~U_{L}$ and the associated expression for
the feedforward filter  ${\bf w}_{k} = {\bf R}^{-1}_{U_{l}}{\bf
p}_{k}$, where $l=1, 2, \ldots, L$. Therefore, the MMSE for user
$k$ is approximately given by:
\begin{equation}
J_{MMSE}  \approx \arg  \min_{1\leq l \leq L} (MSE_{U_{l}})
\end{equation}
where $MSE_{U_{l}} = \sigma^2_{b} - {\bf p}_{k}^{H}{\bf
R}^{-1}_{U_{l}}{\bf p}_{k}$ and the above expression means that
the MMSE attained by user $k$ with the SPA-DF is at least equal to
a standard S-DF (with $L=1$ and approximate MMSE given by (45)).
The approximate MMSE in (47) is also proportional to the number of
undetected users expressed by the covariance matrix ${\bf
R}_{U_{l}}$, but can benefit from different groups of undetected
users, by selecting the undetected group of users that yield
smaller MSE, resulting in better performance. Indeed, the MMSE of
the proposed SPA-DF structure in (47) is  upperbounded by the MMSE
of the standard S-DF detector given through (45).



\section{On the MMSE of Optimal Ordering with S-DF detectors }

Here, we mathematically discuss the MMSE of S-DF detectors with
the optimal ordering algorithm.  If we consider an exhaustive
search over all the possible orderings for an S-DF, we have $K!$
different groups of undetected users or equivalently $K!$ possible
orderings. The optimal ordering S-DF can be seen as a
generalisation of the proposed SPA-DF structure in which the
number of branches is equal to $K!$. Mathematically, for the case
of imperfect decisions we have for the optimal ordering S-DF the
following expression
\begin{equation}
J_{MMSE}  \approx \arg  \min_{1\leq l \leq K!} (J_{MSE,l})
\end{equation}
where
\begin{equation}
\begin{split}
J_{MSE,l} & = \sigma^2_{b} - {\bf p}_{k,l}^{H}{\bf R}^{-1}{\bf
p}_{k} - {\bf f}_{k,l}^{H}{\bf B}^{H}{\bf R}^{-1}{\bf p}_{k} -
{\bf p}_{k}^{H}{\bf R}^{-1}{\bf B}{\bf f}_{k,l}  \\ & \quad -{\bf
f}_{k,l}^{H}{\bf B}^{H}{\bf R}^{-1}{\bf B}{\bf f}_{k,l} + {\bf
f}_{k,l}^{H}{\bf f}_{k,l}
\end{split}
\end{equation}
The expression in (49) is similar in form to the first line of
(42) but depends on the ordering $l$ and the associated feedback
filter ${\bf f}_{k,l}$. In the case of perfect feedback, the
corresponding expression for the feedforward filter is ${\bf
w}_{k} = {\bf R}^{-1}_{U_{l}}{\bf p}_{k}$, where $l=1, 2, \ldots,
K!$ and we have $K!$ different groups of undetected users, namely,
$U_{1},~U_{2}, ~\ldots,~U_{K!}$. Therefore, the MMSE for user $k$
is approximately given by
\begin{equation}
J_{MMSE}  \approx \arg  \min_{1\leq l \leq K!} (MSE_{U_{l}})
\end{equation}
where $MSE_{U_{l}} = \sigma^2_{b} - {\bf p}_{k}^{H}{\bf
R}^{-1}_{U_{l}}{\bf p}_{k}$ and the above expression means that
the MMSE attained by user $k$ with the optimal ordering is at
least equal to a standard S-DF (with $L=1$ and approximate MMSE
given by (45)). The approximate MMSE in (50) is indeed
proportional to the number of undetected users expressed by the
covariance matrix ${\bf R}_{U_{l}}$. The key point is that the
designer searches for all possible groups of undetected users and
selects the one which yields the smallest MSE, resulting in better
performance. The main problem is that as $K$ increases the
complexity becomes prohibitive and its implementation impractical.
\end{appendix}


\begin{thebibliography}{100}
\linespread{1.5}

\bibitem{verdu}
S. Verdu, \textit{ Multiuser Detection}, Cambridge, 1998.

\bibitem{verdu86}
S. Verdu, ``Minimum Probability of Error for Asynchronous Gaussian
Multiple-Access Channels", \textit{IEEE Transactions on
Information Theory}, vol.IT-32, no. 1, pp. 85-96, Janeiro, 1986.

\bibitem{lupas}
R. Lupas and S. Verdu, ``Linear multiuser detectors for
synchronous code-division multiple-access channels," \textit{IEEE
Transactions on Information Theory}, vol. 35, pp. 123–136, Jan.
1989.

\bibitem{falconer}
M. Abdulrahman, A. U.K. Sheikh and D. D. Falconer, ``Decision
Feedback Equalization for CDMA in Indoor Wireless Communications,"
\textit{IEEE Journal on Selected Areas in Communications}, vol 12,
no. 4, May, 1994.

\bibitem{patel}
P. Patel and J. Holtzman, ``Analysis of a Simple Successive
Interference Cancellation Scheme in a DS/CDMA Systems", \textit{
IEEE Journal on Selected Areas in Communications}, vol. 12, n. 5,
June, 1994.

\bibitem{varanasi}
M. K. Varanasi and B. Aazhang, ``Multistage detection in
asynchronous CDMA communications," \textit{ IEEE Trans.
Communications}, vol. 38, no. 4, pp. 509-19, April, 1990.

\bibitem{shamai}
S. Verdu and S. Shamai, ``Spectral efficiency of CDMA with random
spreading," \textit{IEEE Transactions on Information Theory}, vol.
45, pp. 622-640, 1999.

\bibitem{rapajic}
P. B. Rapajic, M. L. Honig and G. K. Woodward, ``Multiuser
decision-feedback detection: Performance bounds and adaptive
algorithms," in \textit{IEEE Int. Symp. on Inform. Theory},
Boston, MA, August, 1998, p. 34.

\bibitem{duel}
A. Duel-Hallen, ``A family of multiuser decision-feedback
detectors for asynchronous CDMA channels," {\it IEEE Transactions
on Communications}, vol. 43, Feb.-Apr. 1995.

\bibitem{varanasi2}
M. K. Varanasi and T. Guess, ``Optimum decision feedback multiuser
equalization with successive decoding achieves the total capacity
of the Gaussian multiple-access channel," in \textit{ Proc. 31st
Asilomar Conf. Signals, Systems and Computers}, Monterey, November
1997, pp. 1405-1409.

\bibitem{varanasi3}
M. K. Varanasi, ``Decision feedback multiuser detection: A
systematic approach," \textit{IEEE Transactions on Information
Theory}, vol. 45, pp. 219-240, January 1999.

\bibitem{luo}
J. Luo, K. R. Pattipati, P. K. Willet and F. Hasegawa, ``Optimal
User Ordering and Time Labeling for Ideal Decision Feedback
Detection in Asynchronous CDMA", \textit{IEEE Transactions on
Communications}, vol. 51, no. 11, November, 2003.

\bibitem{woodward1}
G. Woodward, R. Ratasuk, M. L. Honig and P. Rapajic, ``Multistage
decision-feedback detection for DS-CDMA," {\it Proc. IEEE ICC},
June 1999.

\bibitem{woodward2}
G. Woodward, R. Ratasuk, M. L. Honig and P. Rapajic, ``Minimum
Mean-Squared Error Multiuser Decision-Feedback Detectors for
DS-CDMA," {\it IEEE Transactions on Communications}, vol. 50, no.
12, December, 2002.

\bibitem{woodward3}
M. Honig, G. Woodward and Y. Sun, ``Adaptive Iterative Multiuser
Decision Feedback Detection," {\it IEEE Transactions on Wireless
Communications}, vol. 3, no. 2, March 2004.


\bibitem{barriac}
G. Barriac and U. Madhow, ``PASIC: A New Paradigm for
Low-Complexity Multiuser Detection", \textit{Proc. Conf. on
Inform. Sciences and Systems}, The Johns Hopkins University, March
21-23, 2001.

\bibitem{foerster}
J. Foerster and L. B. Milstein, ``Coding for a coherent DS-CDMA
system employing an MMSE receiver in a Rayleigh fading channel,"
\textit{IEEE Trans. Commun.}, vol. 48, pp. 1909–1918, June 2000.

\bibitem{phoel}
W. G. Phoel and M. L. Honig, ``Performance of coded DS-CDMA with
pilot-assisted channel estimation and linear interference
suppression," \textit{IEEE Trans. Commun.}, vol. 50, pp. 822–832,
May 2002.

\bibitem{alexander1} P. D. Alexander, A. J. Grant, and M. C. Reed,
``Iterative detection in code-division multiple-access with error
control coding," \textit{Eur. Trans. Telecommun.}, vol. 9, pp.
419–425, Sept.-Oct. 1998.

\bibitem{reed}
M. C. Reed, C. B. Schlegel, P. D. Alexander, and J. A.
Asenstorfer, ``Iterative multiuser detection for CDMA with FEC:
Near-single-user performance," \textit{IEEE Trans. Commun.}, vol.
46, pp. 1693–1699, Dec. 1998.

\bibitem{alexander2}
P. D. Alexander, M. C. Reed, J. Asenstorfer, and C. B. Schlegel,
`` Iterative multiuser interference reduction: Turbo CDMA,"
\textit{IEEE Trans. Commun.}, vol. 47, July 1999.

\bibitem{wang}
X. Wang and H. V. Poor, ``Iterative (turbo) soft interference
cancellation and decoding for coded CDMA," \textit{IEEE Trans.
Commun.}, vol. 47, pp. 1046–1061, July 1999.

\bibitem{galmal}
H. E. Galmal and E. Geroniotis, ``Iterative multiuser detection
for coded CDMA signals in AWGN and fading channels," \textit{IEEE
J. Select. Areas Commun.}, vol. 47, pp. 30–41, Jan. 2000.

\bibitem{proakis}
J. G. Proakis, \textit{ Digital Communications, 3rd edition},
Mc-Graw Hill, NY, 1995.

\bibitem{delamaresp}
R. C. de Lamare and R. Sampaio-Neto, ``Adaptive Reduced-Rank MMSE
Filtering with Interpolated FIR Filters and Adaptive
Interpolators", \textit{IEEE Signal Processing Letters}, vol. 12,
no. 3, March, 2005.

\bibitem{vogelbruch}
F. Vogelbruch and S. Haar, ``Low complexity turbo equalization
based on soft feedback interference cancellation", \textit{IEEE
Communications Letters}, vol. 9, no. 7, July 2005.




\end{thebibliography}
\end{document}